\documentclass{article}

\usepackage{graphicx}
\begin{document}

\title{Renormalization in Coulomb gauge QCD}

\author{A. Andra\v si  \footnote{aandrasi@rudjer.irb.hr} \\
{\it `Rudjer Bo\v skovi\' c' Institute, Zagreb, Croatia} \\ \\
  \and
John C. Taylor\footnote{J.C.Taylor@damtp.cam.ac.uk} \\
{\it Department of Applied Mathematics and Theoretical Physics,}\\
{\it University of Cambridge, UK} }

\date{27 Oct. 2010}
\maketitle

\begin{abstract}

{\noindent In the Coulomb gauge of QCD, the Hamiltonian 
 contains a non-linear Christ-Lee term,
which may alternatively be derived from a careful treatment of ambiguous Feynman
integrals at 2-loop order. We investigate how and if UV divergences from higher order graphs
can be consistently absorbed by renormalization of the Christ-Lee term. We find that
they cannot.}\\

\noindent{Pacs numbers: 11.15.Bt; 11.10.Gh} \\

\noindent{Keywords: Coulomb gauge; Renormalization; QCD}
\end{abstract}

\vfill\newpage

\section{Introduction}

The Coulomb gauge in QCD has some attractive properties. It is the only gauge which is explicitly
unitary, all the state vectors being physical (with transverse gluons) with positive norm. (Axial gauges suffer from
ambiguous denominators $1/n.k$ in Feynman propagators.) The Coulomb gauge
has been used in lattice simulations, see for instance\cite{Cucchieri}.
  
  Nevertheless, the Coulomb gauge is not straightforward. First, in individual Feynman
  diagrams, even at 1-loop order, there are linear ``energy divergences" of the form
  \begin{equation}
  \label{1}
  \int dk_0 F
  \end{equation}
  where $F$ is independent of $k_0$. This problem is cured by going to the Hamiltonian,
  phase-space, formalism, in which the conjugate field $E_i^a$ to $A_i^a$ is introduced. Even then,
  introducing quark loops brings back energy divergences in individual graphs
  which must be cancelled by combining graphs \cite{AT}.
  
  In the Hamiltonian formalism, at 1-loop level, there are formally divergent integrals
  of the form (we use $P$ for the spacial part of the 4-vector $p$)
  \begin{equation}
  \label{2}
  \int dp_0 {p_0\over p_0^2-P^2} F
  \end{equation}
  where again $F$ is independent of $p_0$. It is natural to take (2) to be zero. This can be justified
  by taking the Coulomb gauge to be the limit, when a certain parameter tends to zero, of a gauge interpolating between 
  the Feynman gauge and the Coulomb gauge \cite{cheng} \cite{doust}.
  
  To 2-loop order there are more subtle difficulties, in the appearance of
  non-convergent
   integrals of
  the form
  \begin{equation}
  \label{3}
  \int dp_0dq_0 {p_0 \over p_0^2-P^2}{q_0\over q_0^2-Q^2}F(P,Q,....)
  \end{equation}
  (where a Feynman $i\eta$ is understood
  in the denominators).
  It has been shown \cite{mohapatra} \cite{doust} that these divergences are resolved
  when suitable sets of graphs are added. This is achieved partly
   by the use of the 
  identity
  \begin{eqnarray}
  \label{4}
  \int dp_0dq_0dr_o\delta(p_0+q_0+r_0)\left[{p_o\over p_0^2-P^2}{q_0\over q_0^2-Q^2}+
  {q_0\over q_0^2-Q^2}{r_0\over r_0^2-R^2}+{r_0\over r_0^2-R^2}{p_0\over p_0^2-P^2}\right] \nonumber \\
  =-\pi^2
  \end{eqnarray}
  though again an interpolating gauge is necessary for a complete rigorous treatment.
  We shall call graphs which contain the integral (3)
   {\it A-graphs} ({\it A } for ``ambiguous")
  and non-convergent integrals like (3) {\it A-integrals}.
  Another rule about the {\it A-integrals} is that integrals like the square of (2), that is of the form
  \begin{equation}
  \label{5}
  \int dp_0 {{p_0}\over{p^2_0-P^2}}F(P)
  \int dq_0{{q_0}\over{q^2_0-Q^2}}G(Q)
   \end{equation}
  are zero. This is consistent with (2), but the rule can again be justified rigorously \cite{doust} by 
  using an interpolating gauge. We call this rule "factorization".
  
  Previous to this work, an equivalent result has been derived by Christ and Lee \cite{christlee}.
  They noted an operator ordering ambiguity in the non-local Coulomb interaction in the
  Coulomb gauge Hamiltonian. They resolved this by the addition of extra operators
  to the Hamiltonian, at the same time defining the Feynman integration so that
  integrals like (3) are zero. (Though this definition must be used with caution in view of the
  identity (4).)

  All the above work is concerned with the energy integrals, with the spatial momenta temporarily
  held fixed. 
  When this is done, it has been shown \cite{mohapatra} \cite{doust} that non-convergent {\it A-integrals} appear
  at 2-loop order only, not at 1-loop and not at 3-loop and higher.
   Problems might arise when the above energy divergences are considered along with 
  ordinary UV divergences and the necessity of renormalization \cite{doust} \cite{taylor}.
  The standard method of renormalization demands that UV divergent sub-graphs are
  computed and renormalized before insertion in the main graph. This order of integration
  might conflict with the use of (4), which requires two energy integrals to be done
  before either of the corresponding integrals over spatial momenta. However,  consider the
  sub-graphs  of an {\it A-graph}. Such a sub-graph contains (2) and so is zero. Therefore  there is no
  UV divergence in the sub-graph and no renormalization is needed.
  This is consistent, since a Feynman integral of the form (with dimensional regularization)
  \begin{equation}
  \label{6}
  \int d^{3-\epsilon}Pdq_0dp_0\frac{p_0}{p^2-P^2}\frac{q_0}{q_0^2-Q^2}H(P,Q)
  \end{equation}
  never has a pole ${1\over{\epsilon}}$,
  where $\epsilon=4-n$ and $n$ is the number of space-time dimensions.
  
  In this paper, we study another such possible conflict: the insertion of UV divergent sub-graphs
  into {\it A-graphs}. For simplicity, we choose the sub-graphs to
  be quark loops (which would dominate for large $N_f$, the number of families).
  We are not permitted to perform all three energy integrals first, with all
  spatial momenta held fixed: renormalization demands that we do all the
  integrals, energy and spatial, in the UV divergent sub-graph first.
  Thus we are concerned with integrals of the form

   \begin{equation}
   \label{7}
   \int dp_0dq_0ds_0 d^{3-\epsilon} 
   S J(p_0,q_0,P,Q;s_0,S)
   \end{equation}
  where the $q$-intergration is an UV divergent sub-graph (a fermion loop). We hold $P$ and $Q$
  fixed, but we have to do both the $s_0$ and the $S$ integrations first because a renormaization
  subtraction must be made before doing other integrals.
  So we require the high-energy behaviour of the subgraphs (2, 3, and 4-gluon
  diagrams)
  in order to study the convergence of the remaining two energy integrals.
  We can obtain this high-energy behaviour from the Ward identities
   obeyed by the quark loops which
   determine all the high energy behaviours
  in terms of one function, the gluon self-energy $S(p)$. When an attempt is made to
  use the identity (4), this function $S$ appears in various places. The question is whether
  these extra insertions spoil the identity. We find that they do. We conclude that the attempt
  to combine UV renormalization with the control of energy divergences leads to trouble.
  
  Another way to regard this problem is to ask what are the fields and coupling constants
  in the Christ-Lee operator. Are they bare or renormalized quantities? Since Christ and Lee
  derived this from consideration of operation ordering of the original
  Hamiltonian, it seems they must be bare quantities. Then the question is how to
  re-express the operator in terms of useful, renormalized quantities.

  In section 3, we derive the high-energy behaviour of the quark sub-graph loops by using the Ward identites,
  in section 4 we give the results for individual graphs and in section 5 there are conclusions.
  
  \section{Notation and graphs}
  
  We use the same notation and graphical conventions as in \cite{AT2}.  As well as lines for Coulomb
  interactions and for transverse gluon propagators, there are lines corresponding to transitions
  to $E_i^a$.

  The relevant 2-loop graphs contain exactly two transverse propagators together with three
  Coulomb lines. They contain no  vertex where three transverse gluons meet.

  \begin{figure}
  \begin{center}
   \includegraphics[width=7.5cm]{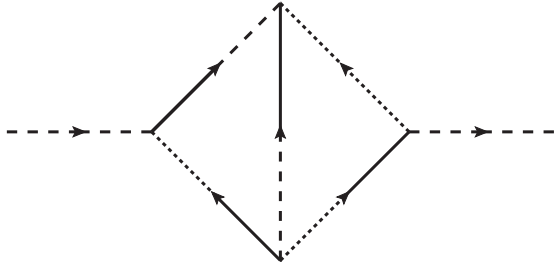}%
  \caption{The example of the 2-loop graphs, graph 1B. Continuous lines represent ${\bf E} $,
   dashed lines $ {\bf A} $ and dotted lines $ {\bf A_0 }$.}
  \end{center}
  \end{figure}
  
  \section{The high-energy limit of the quark loops}
  
  Let $t^a$ be the colour matrices in the quark representation, with
  \begin{equation}
  \label{8}
 \mbox{tr}(t^a t^b)=C_q\delta^{ab}.
  \end{equation}
  
  The gluon self-energy from the quark loop is
  \begin{equation}
  \label{9}
 \mbox{tr}(t^at^b)S_{\mu_1\mu_2}(p)=g^2C_q\delta^{ab}(p_{\mu_1}p_{\mu_2}-p^2\delta_{\mu_1\mu_2})S(p^2),
 \end{equation}
 where
 \begin{equation}
 \label{10}
 S(p^2)=8i\pi^{2-{{\epsilon}\over 2}}\Gamma({{\epsilon}\over 2})
  {{\Gamma^2(2-{{\epsilon}\over 2})}\over{\Gamma(4-\epsilon)}}
  \left[(-p^2-i\eta)^{-{{\epsilon}\over2}}-(\mu^2)^{-{{\epsilon}\over2}}\right],
  \end{equation}
 where a renormalization subtraction at a mass $ \mu $ has been made.
 
 The quark triangle is
 \begin{equation}
 \label{11}
 \mbox{tr}(t^a[t^b,t^c])V_{\mu_1\mu_2\mu_3}(p_1,p_2,p_3)\delta(p_1+p_2+p_3)
 \end{equation}
 where $V$ is totally anti-symmetric under permutations of $1,2,3$.
  
  The quark square is
  \begin{equation}
  \label{12}
  \mbox{tr}(t^at^bt^ct^d+t^dt^ct^bt^a)W_{\mu_1\mu_2\mu_3\mu_4}(p_1,p_2,p_3,p_4)\delta(p_1+p_2+p_3+p_4),
  \end{equation}
  where $W$ has cyclic symmetry in $1,2,3,4$ and symmetry under $1,2,3,4\rightarrow 4,3,2,1$.
  Because of these symmetries, there are in general three independent tensors $W$.
  But in the present case, the high-energy limits of the $W'$s are independent of the quark
  representation (apart from the overall factor $C_q$), and as a consequence
   there is an additional relation
  \begin{equation}
  \label{13}
  W_{\mu_1\mu_2\mu_3\mu_4}W(k_1,k_2,k_3,k_4)+(1,2,3\rightarrow 3,1,2)+(1,2,3\rightarrow 2,3,1)=0.
  \end{equation}
  The Ward identities connecting the quark loops are
  \begin{equation}
  \label{14}
  k_{30}V_{000}-K_{3i}V_{00i}=S_{00}(k_1)-S_{00}(k_2)=K^2_1 S(k_1)-K^2_2 S(k_2),
  \end{equation}
  \begin{equation}
  \label{15}
  k_{20}V_{00i}-K_{2j}V_{0ji}=S_{0i}(k_3)-S_{0i}(k_1)=k_{30}K_{3i}S(k_3)-k_{10}K_{1i}S(k_1),
  \end{equation}
  \begin{equation}
  \label{16}
  k_{10}V_{0ij}-K_{1l}V_{lij}=S_{ij}(k_2)-S_{ij}(k_3)= \delta_{ij}[k^2_{20}S(k_2)-k^2_{30}S(k_3)],
  \end{equation}
  \begin{equation}
  \label{17}
  k_{10}W_{00kl}-K_{1i}W_{i0kl}=V_{0kl}(k_1+k_2,k_3,k_4)-V_{0kl}(k_1,k_2,k_3+k_4),
  \end{equation}
  \begin{eqnarray}
  \label{18}
  k_{10}W_{0i0l}(k_1,k_2,k_3,k_4)-K_{1m}W_{mi0l}(k_1,k_2,k_3,k_4) \nonumber \\  
   = V_{i0l}(k_1+k_2,k_3,k_4)-V_{i0l}(k_2,k_3,k_1+k_4).
  \end{eqnarray}

  These may be solved in the limit where the time components of the momenta are much
  larger than the space components, to give
  \begin{equation}
  \label{19}
  V_{0ij}(k_1, k_2, k_3)\approx k_{10}^{-1}\delta_{ij}\left[k^2_{20}S(k_2)-k^2_{30}S(k_3)\right],
  \end{equation}
  \begin{equation}
  \label{20}
  V_{00i}(k_1, k_2, k_3)\approx {{K_{2i}}\over{k_{10}}}\left[k_{20}S(k_2)+k_{30}S(k_3)\right]-{{K_{1i}}\over{k_{20}}}\left[k_{30}S(k_3)+k_{10}S(k_1)\right],
  \end{equation}
  \begin{eqnarray}
  \label{21}
  V_{000}(k_1, k_2, k_3)& \approx & -{{K_2\cdot K_3}\over{k_{10}}}\left[S(k_2)-S(k_3)\right]
                             -{{K_3\cdot K_1}\over{k_{20}}}\left[S(k_3)-S(k_1)\right] \nonumber \\
               & &               -{{K_1\cdot K_2}\over{k_{30}}}\left[S(k_1)-S(k_2)\right].
  \end{eqnarray}
  We use notation $ k_{12}=k_1+k_2$ etc,
  \begin{equation}
  \label{22}
  W_{00ml}(k_1, k_2, k_3, k_4)\approx -\delta_{ml}\left[ {{k^2_{30}S(k_3)}\over{k_{20}k_{120}}}+{{k^2_{40}S(k_4)}\over{k_{10}k_{120}}}
                 -{{k^2_{140}S(k_1+k_4)}\over{k_{10}k_{20}}}\right],
  \end{equation}
  \begin{eqnarray}
  \label{23}
  W_{0i0l}(k_1, k_2, k_3, k_4)& \approx & {1\over{k_{10}k_{30}}}\delta_{il}\left[k^2_{40}S(k_4)+k^2_{20}S(k_2)- (k_1+k_2)^2_0S(k_1+k_2) \right.  \nonumber  \\
  & &  \left.   -(k_1+k_4)^2_0S(k_1+k_4)\right], 
  \end{eqnarray}
  \begin{eqnarray}
  \label{24}
  W_{000i}(k_1, k_2, k_3, k_4)\approx {{K_{2i}}\over{k_{10}k_{30}}}\left[k_{340}S(k_3+k_4)+k_{140}S(k_1+k_4)+k_{20}S(k_2)-k_{40}S(k_4)\right]\nonumber \\
  +K_{1i}\left[-{{k_{340}}\over{k_{20}k_{30}}}S(k_3+k_4)-{{k_{10}}\over{k_{20}k_{230}}}S(k_1)+{{k_{40}}\over{k_{30}k_{230}}}S(k_4)\right]\nonumber \\  
  +K_{3i}\left[-{{k_{140}}\over{k_{20}k_{10}}}S(k_1+k_4)-{{k_{30}}\over{k_{20}k_{120}}}S(k_3)+{{k_{40}}\over{k_{10}k_{120}}}S(k_4)\right],                
   \end{eqnarray}
  \begin{eqnarray}
  \label{25}
  W_{0000}(k_1, k_2, k_3, k_4)& \approx & {{K_1\cdot K_3}\over{k_{20}k_{40}}}\left[S(k_2+k_3)+S(k_3+k_4)-S(k_1)-S(k_3)\right]\nonumber \\
   & & -{{K_2\cdot K_4}\over{k_{10}k_{30}}}\left[S(k_1+k_4)+S(k_3+k_4)-S(k_2)-S(k_4)\right]\nonumber \\
  & & +(K_1\cdot K_2)\left[{{S(k_1)}\over{k_{40}k_{340}}}+{{S(k_2)}\over{k_{30}k_{340}}}-{{S(k_1+k_4)}\over{k_{30}k_{40}}}\right] \nonumber \\
  & & +(K_3\cdot K_4)\left[{{S(k_3)}\over{k_{20}k_{120}}}+{{S(k_4)}\over{k_{10}k_{120}}}-{{S(k_1+k_4)}\over{k_{10}k_{20}}}\right]\nonumber \\
  & & +(K_2\cdot K_3)\left[{{S(k_3)}\over{k_{40}k_{140}}}+{{S(k_2)}\over{k_{10}k_{140}}}-{{S(k_3+k_4)}\over{k_{40}k_{10}}}\right]\nonumber \\
  & & +(K_4\cdot K_1)\left[{{S(k_4)}\over{k_{30}k_{230}}}+{{S(k_1)}\over{k_{20}k_{230}}}-{{S(k_1+k_2)}\over{k_{20}k_{30}}}\right].
  \end{eqnarray}
  
  The other components of $V$ and $W$ are negligible in this limit.
  Note that all the equations above apply only to the region where all the time components are much larger than any space
  component. So these equations are useful for checking convergence but not for finding the actual values of integrals.
  All subsequent equations are to be understood in the same way.
  
  \section{The gluon graphs with quark insertions}
  
  To take the simplest case, we look at graphs with just two external transverse
  gluon lines, with momenta $k$ (in the Coulomb gauge, both real and virtual gluons are transverse).
  
  We wish to test whether the ambiguous integrals like (2) combine into the unambiguous
  combinations (3) when quark loops insertions are present. We expect this to happen
  with the spatial parts of the gluon momenta held fixed, and it should therefore happen
  as an identity in the spatial momenta. With the basic set of independent invariant functions of $ P$, $Q$, $K$,
  \begin{equation}
  \label{26}
  K^2,~~~ P^2,~~~ Q^2,~~~ P'^2=(P-K)^2,~~~ Q'^2=(Q-K)^2,~~~ R'^2=(K-P-Q)^2
  \end{equation}
  we choose an algebraically independent set of functions of these spatial momenta.
  We have the following classes of invariant functions
  \begin{equation}
  \label{27}
  (a)~~~~~ {{K^2}\over{P^2P'^2Q^2Q'^2}}~~~~~~~~~~~~~~
  \end{equation}
   etc., i.e. with $K^2$ in the numerator and 4 different denominators (there are 5 such functions),
  \begin{equation}
  \label{28}
   (b)~~~~~{1\over{P^2Q^2R'^2}}~~~~~~~~~~~~~~~~~
  \end{equation}
  etc. with 3 denominators (there are 10 such functions),
  \begin{equation}
  \label{29}
   (c)~~~~~{{P^2}\over{Q^2Q'^2P'^2R'^2}}~~~~~~~~~~~~
  \end{equation}
   etc., that is one numerator (not $K^2$) and 4 other denominators (there are 5 such functions).
   Out of these 20 independent functions, we choose (27) in order to make the test.
   The functions in (a), (b) and (c) are sufficient for the graphs in fig.3 till fig.26.
   There is another class of diagrams,
   containing an internal self-energy part, of which one example is shown in fig.2. 
   No graph of this second type can
   contribute to the invariant in (27), no matter how the variables $P$ and $Q$ are defined; so we need
   not consider any graph of this type.

   \begin{figure}
   \begin{center}
    \includegraphics[width=7.5cm]{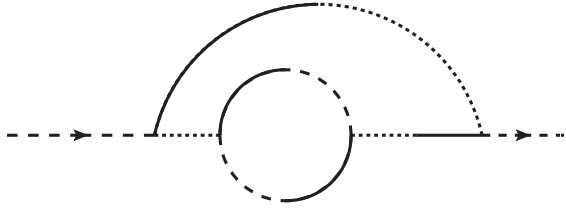}%
   \caption{The example of the other class of diagrams which cannot contribute to the invariant in (27).}
   \end{center}
   \end{figure}
  
   The relevant graphs which contribute to
   $K^2$ are shown in fig.3 till fig.26. 
   We use the notation
   \begin{equation}
   \label{30}
    \alpha_{ab}={1\over 4}g^6(2\pi)^{-8}C^2_q T(R)\delta_{ab}K^2\int d^4 p \int d^4 q{1\over{P^2P'^2Q^2Q'^2}}.
    \end{equation}
   The graphs are grouped into sets for which large cancellations of linear energy
    divergences occurr.
   The notation, for example $ 1B(t, ij0) $ describes the fermion loop inserted on top of the original graph
   $ 1B $ connected with two transverse (indices i, j) and one Coulomb line to the rest of the graph.
   There are several original two-loop graphs (e.g. $ 1B, 2B, 3B $) which differ only in the positions of 
   Coulomb and transverse lines. The momenta are defined as $ p'=p-k,$ $q'=q-k,$ $r'=k-p-q.$ 
   We list the final results for the graphs.

   \subsection {Set 1}
   In this subsection we consider the graphs shown in fig.3 and fig.4, between which large cancellations of linear
   energy divergneces take place.

   The sum of 3 graphs (one representative graph is shown in fig.3) is

   \begin{equation}
   \label{31}
    1B(t, ij0)+2B(t, ij0)+3B(t, ij0)=\alpha_{ab}P_i(Q+2Q')_j{1\over{r'^2_0 p_0q'_0}}[p^2_0 S(p) -r'^2_0 S(r')].
   \end{equation}
   
   \begin{figure}
   \begin{center}
    \includegraphics[width=7.5cm]{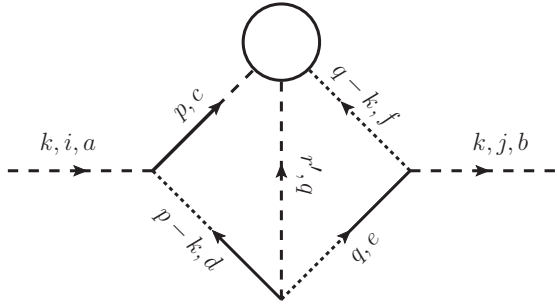}%
   \caption{Graph $1B(t, ij0)$. There are two other graphs in which the pure Coulomb line is p' or q
      rather than q', giving $ 2B(t, ij0) $ and $ 3B(t, ij0) $ respectively. 
      The arrow indicates the sense of momentum flow.}
   \end{center}
   \end{figure}

   \begin{figure}
    \begin{center}
     \includegraphics[width=7.5cm]{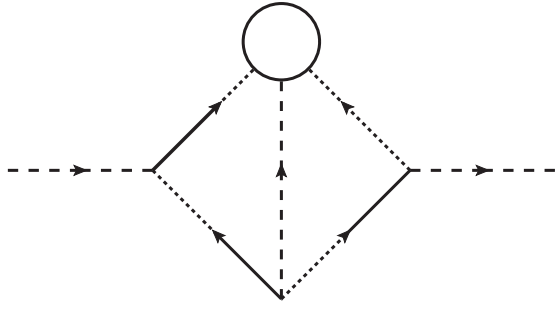}%
    \caption{Graph $1B(t, 0j0)$. There are 3 other graphs obtained by moving the Coulomb line
        around the gluon loop.}
    \end{center}
    \end{figure}
    
   We have 3 graphs obtained from graphs in (30) by rotating the internal lines about 
   the vertical axis (keeping the external gluons fixed) giving

   \begin{equation}
   \label{32}
    1B(t, 0jk)+2B(t, 0jk)+3B(t, 0jk)=\alpha_{ab}(2P+P')_iQ'_j{1\over{r'^2_0p_0q'_0}}[q'^2_0S(q')-r'^2_0S(r')].
   \end{equation}

   There are 4 graphs (the example $ 1B(t, 0j0) $ is shown in fig.4)
   where the fermion loop is connected to the rest of the graph with two Coulomb lines.
   Their sum amounts to
    \begin{eqnarray}
    \label{33}
    1B(t, 0j0)+2B(t, 0j0)+3B(t, 0j0)+4B(t, 0j0)=-\alpha_{ab}(P_iQ_j+2P_iQ'_j+P'_iQ'_j)\nonumber \\
    \times {1\over{r'^2_0}}\{{1\over{q'_0}}[p_0S(p)+r'_0S(r')]+{1\over{p_0}}[r'_0S(r')+q'_0S(q')]\}. 
    \end{eqnarray}

    We see that individual terms in this set contain linear energy divergences, for example $ {1\over{r'^2_0}}.$
    But in the sum of eqs.(31), (32) and (33) large cancellations of linear divergences occurr, giving for the
    first set

    \begin{equation}
    \label{34}
    Set 1=-\alpha_{ab}\{{1\over{r'^2_0}}[{{q'_0}\over{p_0}}P_iQ_jS(q')+{{p_0}\over{q'_0}}P'_iQ'_jS(p)]+{2\over{p_0q'_0}}P_iQ'_jS(r')\}.
    \end{equation}

    \subsection {Set 2}

    In this set we treat graphs with fermion loop on the left side, connected to the incoming gluon.
    There are 3 distinct graphs like the graph $ 4B(l, ij0)$ shawn in Fig.5. Their sum is
  
    \begin{equation}
    \label{35}
    4B(l, ij0)+5B(l, ij0)+6B(l, ij0)=-\alpha_{ab}P_i(Q+2Q')_j{1\over{r'^2_0}}S(p).
    \end{equation}

    \begin{figure}
    \begin{center}
    \includegraphics[width=7.5cm]{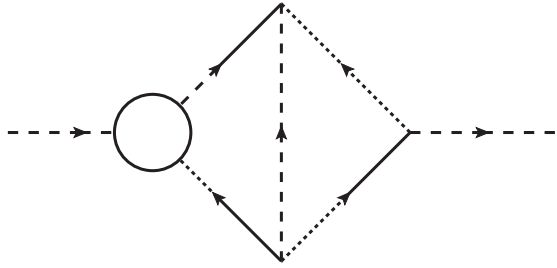}%
    \caption{Graph $4B(l, ij0)$. There are two more distinct graphs in this class. }
    \end{center}
    \end{figure}

    We have two graphs where the fermion loop connects through two Coulomb lines.
    
    \begin{equation}
    \label{36}
    1B(l, 0i0)=\alpha_{ab}{1\over{r'^2_0}}Q'_j[P_iS(p)+P'_iS(p')],
    \end{equation}

    \begin{equation}
    \label{37}
    1B'(l, 0i0)=\alpha_{ab}{1\over{r'^2_0}}Q_j[P_iS(p)+P'_iS(p')].
    \end{equation}
    
   In the $ S(p') $ terms in (36) and (37) we make the change of variables of integration
    $ p \leftrightarrow -p',
   q \leftrightarrow-q', r \leftrightarrow -r, k \leftrightarrow k, (i,j \leftrightarrow i, j) $
   and obtain

   \begin{equation}
   \label{38}
    1B(l, 0i0)+1B'(l, 0i0)= \alpha_{ab}{1\over{r'^2_0}}(Q+Q')_j \times 2P_iS(p).
   \end{equation}

   \begin{figure}
   \begin{center}
    \includegraphics[width=7.5cm]{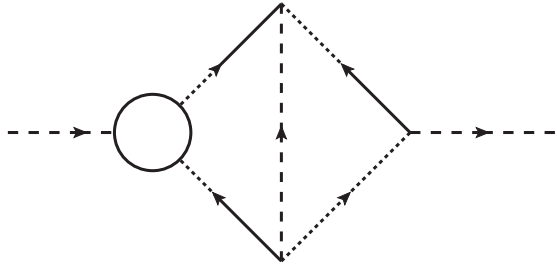}%
   \caption{Graph $ 1B(l, 0i0) $. The other graph $ 1B'(l, 0i0) $ has $p$ and $q$ lines interchanged.}
   \end{center}
   \end{figure}

    Again large cancellations of linear divergences occurr in the sum of (38) and (35) giving

  \begin{equation}
  \label{39}
    Set 2=\alpha_{ab}{1\over{r'^2_0}}P_iQ_jS(p).
  \end{equation}

  \subsection {Set 3}

   $ Set 3 $ consists of 5 graphs which are rotations about the vertical axis (keeping the external gluons fixed)
   of the graphs contained in $ Set 2. $

  \begin{equation}
  \label{40}
  Set 3=\alpha_{ab}{1\over{r'^2_0}}P'_iQ'_jS(q').
  \end{equation}

  In the limit of large linear divergences (i.e. $ r'_0=-p_0-q'_0 \approx 0 $) the sum of the first three sets of diagrams gives
  
  \begin{equation}
  \label{41}
  Set 1+Set 2+Set 3=\alpha_{ab}\{{1\over{r'^2_0}}(P_iQ_j+P'_iQ'_j)[S(p)+S(q')]-{2\over{p_0q'_0}}P_iQ'_jS(r') \}.
  \end{equation}

  \subsection{Set 4}

  This set contains four self-energy graphs. Their sum cancels the linear divergence in (41).

  \begin{equation}
  \label{42}
   Set 4=-\alpha_{ab}{1\over{r'^2_0}}(P_iQ_j+P'_iQ'_j)[S(p)+S(q')]
  \end{equation}

  \begin{figure}
  \begin{center}
   \includegraphics[width=7.5cm]{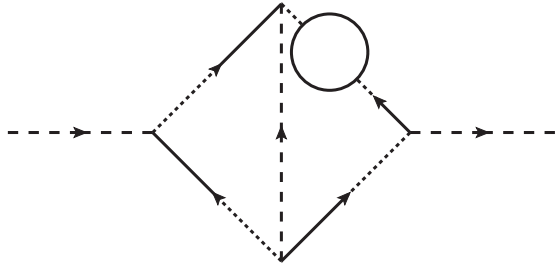}%
  \caption{Graph $SE1$ which represents the four graphs in $ Set 4 $. We insert the fermion loop on $ q'$ or $ p $ line. 
  All other graphs which could be drawn
    in this set are rotations about the horizontal axis and so identical to the first four graphs. }
  \end{center}
  \end{figure}

  \subsection{Set 5}

  There are four graphs in this set. We give the results and respective figures.
 \begin{equation}
 \label{43}
  1B(b, 0i0)=-\alpha_{ab}P_iQ_j{1\over{p_0r'_0}}\{{1\over{p'_0}}[q_0S(q)+r'_0S(r')]+{1\over{q_0}}[p'_0S(p')+r'_0S(r')]\},
 \end{equation}

 \begin{equation}
 \label{44}
 1\widetilde {B}(t, 0i0)=-\alpha_{ab}P_iQ_j{1\over{q_0r'_0}}\{{1\over{q'_0}}[p_0S(p)+r'_0S(r')]+{1\over{p_0}}[q'_0S(q')+r'_0S(r')]\},
 \end{equation}

 \begin{figure}
 \begin{center}
  \includegraphics[width=7.5cm]{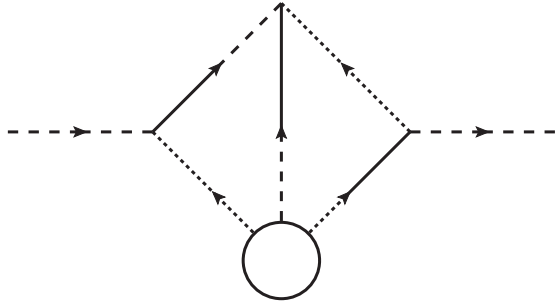}%
  \caption{Graph $ 1B(b, 0i0). $ The fermion loop on bottom is connected with two Coulomb lines to the rest of the graph. }
 \end{center}
 \end{figure}

 \begin{figure}
 \begin{center}
 \includegraphics[width=7.5cm]{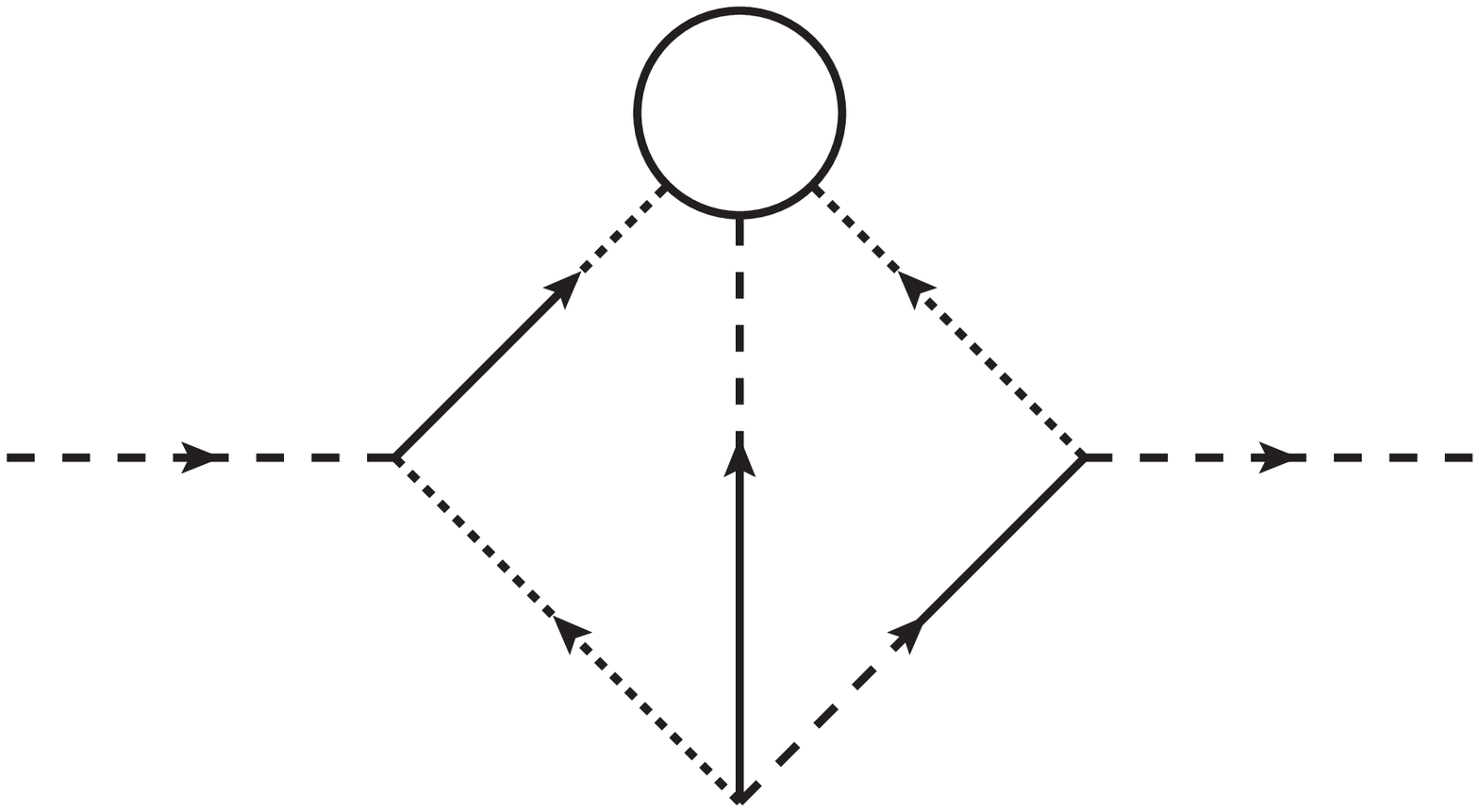}%
 \caption{Graph $1\widetilde {B}(t, 0i0)$}
 \end{center}
 \end{figure}

 \begin{equation}
 \label{45}
 1B(b, 0ij)=-\alpha_{ab}P_iQ_j{1\over{q_0r'_0p_0p'_0}}[r'^2_0S(r')-q^2_0S(q)],
 \end{equation}

 \begin{figure}
 \begin{center}
  \includegraphics[width=7.5cm]{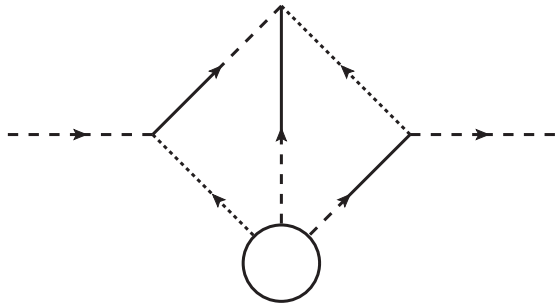}%
 \caption{Graph $ 1B(b, 0ij)$ }
 \end{center}
 \end{figure}

 \begin{equation}
 \label{46}
 1\widetilde {B}(t, 0ij)=-\alpha_{ab}P_iQ_j{1\over{p_0r'_0q_0q'_0}}[r'^2_0S(r')-p^2_0S(p)].
 \end{equation}

 \begin{figure}
 \begin{center}
  \includegraphics[width=7.5cm]{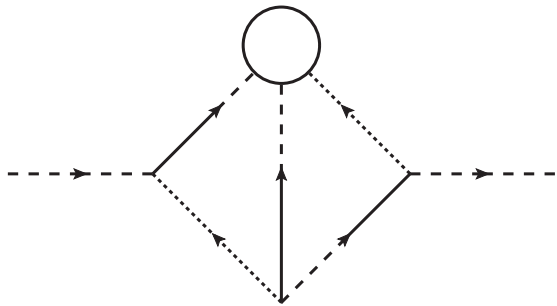}%
 \caption{Graph $1\widetilde {B}(t, 0ij) $}
 \end{center}
 \end{figure}
 
 Again large cancellations of linear divergences occurr in the sum of graphs in $ Set 5 $ giving

 \begin{equation}
 \label{47}
 Set 5=-\alpha_{ab}{1\over{p_0q_0r'_0}}[p'^2_0S(p')+q'_0S(q')]
 \end{equation}

\subsection {Set 6}

 There are only two graphs in this set.

 \begin{equation}
 \label{48}
 2B(r, 0ij)=\alpha_{ab}P'_iQ'_j{1\over{p'_0r'_0q_0q'_0}}[k^2_0S(k)-q'^2_0S(q')].
 \end{equation}

 \begin{figure}
 \begin{center}
  \includegraphics[width=7.5cm]{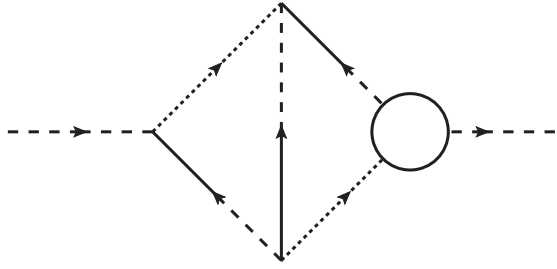}%
  \caption{Graph $ 2B(r, 0ij) $}
 \end{center}
 \end{figure}

 \begin{equation}
 \label{49}
 2\widetilde {B}(l, 0ij)=\alpha_{ab}P'_iQ'_j{1\over{q'_0r'_0p_0p'_0}}[k^2_0S(k)-p'^2_0S(p')].
 \end{equation}

 \begin{figure}
 \begin{center}
  \includegraphics[width=7.5cm]{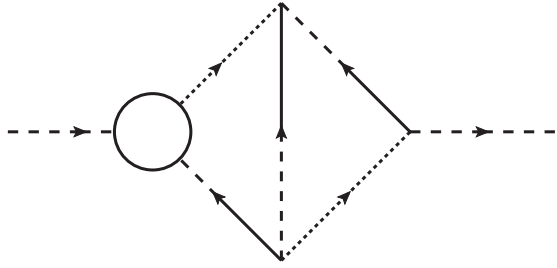}%
 \caption{Graph $2\widetilde {B}(l, 0ij) $ }
 \end{center}
 \end{figure}

In the high-energy limit we ignore terms $ k_0S(k) $ and take $ {{q'_0}\over{q_0}} \approx 1 $, $ {{p'_0}\over {p_0}}\approx 1 $,
so the sum of two graphs in this set amounts to

 \begin{equation}
 \label{50}
 Set 6=-\alpha_{ab}P'_iQ'_j{1\over{p'_0q'_0r'_0}}[p'_0S(p')+q'_0S(q')].
 \end{equation}

 We also have a graph $ 1B(b, 00j) $ which has no $ K^2 $ contribution.
 
 \begin{figure}
 \begin{center}
  \includegraphics[width=7.5cm]{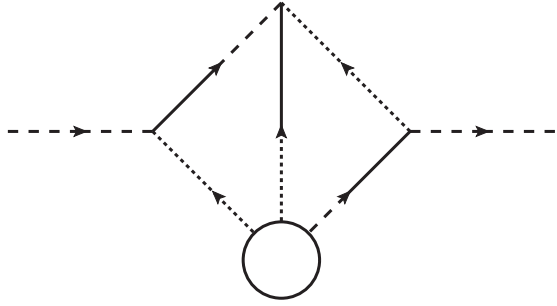}%
 \caption{Graph $ 1B(b, 00j).$ This graph has no $ K^2 $ contribution.}
 \end{center}
 \end{figure}

 \subsection {Set 7}

 There are 3 graphs with a left vertex part
 insertion and 15 graphs with 
 a self-energy insertion in this set.

 The first three give
 \begin{equation}
 \label{51}
 1B(l, ij0)+2B(l, ij0)+3B(l, ij0)=\alpha_{ab}P_i(Q+2Q')_j{1\over{p_0r'_0}}S(p).
 \end{equation}

 The graphs with self energy insertions give altogether
 \begin{equation}
 \label{52}
 SE1B+SE2B+SE3B=-\alpha_{ab}P_i(Q+2Q')_j{1\over{p_0r'_0}}[S(p)+S(p')+S(r')+S(q)+S(q')].
 \end{equation}
 
  \begin{figure}
 \begin{center}
  \includegraphics[width=7.5cm]{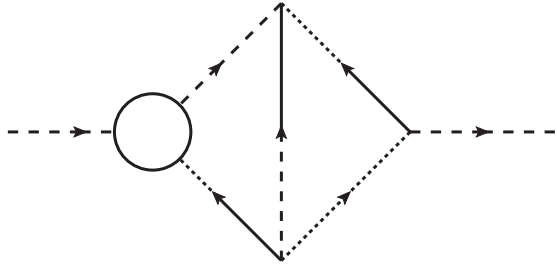}%
 \caption{Graph $ 3B(l, ij0).$ There are two more distinct graphs with left insertion of the fermion loop.}
 \end{center}
 \end{figure}

 \begin{figure}
 \begin{center}
  \includegraphics[width=7.5cm]{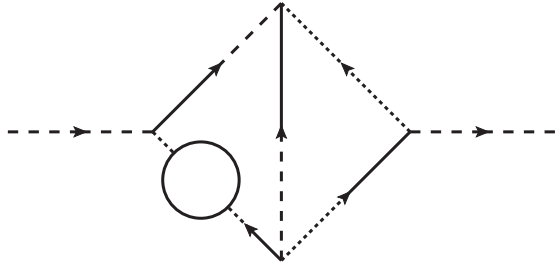}%
 \caption{Graph $ SE1B $. There are 15 graphs with self-energy insertions where the middle line is the $ A_iE_j $-transition.
         We draw just one representative of these graphs. }
 \end{center}
 \end{figure}

 The sum of (52) and (51) amounts to

 \begin{equation}
 \label{53}
  Set 7=-\alpha_{ab}P_i(Q+2Q')_j{1\over{p_0r'_0}}[S(p')+S(r')+S(q)+S(q')].
 \end{equation}

 \subsection {Set 8}

 $ Set 8 $ consists of graphs which are rotations of graphs in $ Set 7 $ about the vertical axis and 
 the change of variables of integration $ p \leftrightarrow -p', q\leftrightarrow -q', k\leftrightarrow k, (i,j\leftrightarrow i,j)$
 \begin{equation}
 \label{54}
 Set 8=-\alpha_{ab}(P+2P')_iQ_j{1\over{q_0r'_0}}[S(q')+S(r')+S(p)+S(p')].
 \end{equation}

 \subsection {Set 9}

  There are two graphs in this set.

 \begin{equation}
 \label{55}
  1\widetilde {B}(r, 0i0)=\alpha_{ab}{1\over{p_0r'_0}}P_i[Q_jS(q)+Q'_jS(q')].
 \end{equation}

 \begin{equation}
 \label{56}
 1\widetilde {B}(l, 0i0)=\alpha_{ab}{1\over{q_0r'_0}}Q_j[P_iS(p)+P'_iS(p')].
 \end{equation}
 
  \begin{figure}
 \begin{center}
  \includegraphics[width=7.5cm]{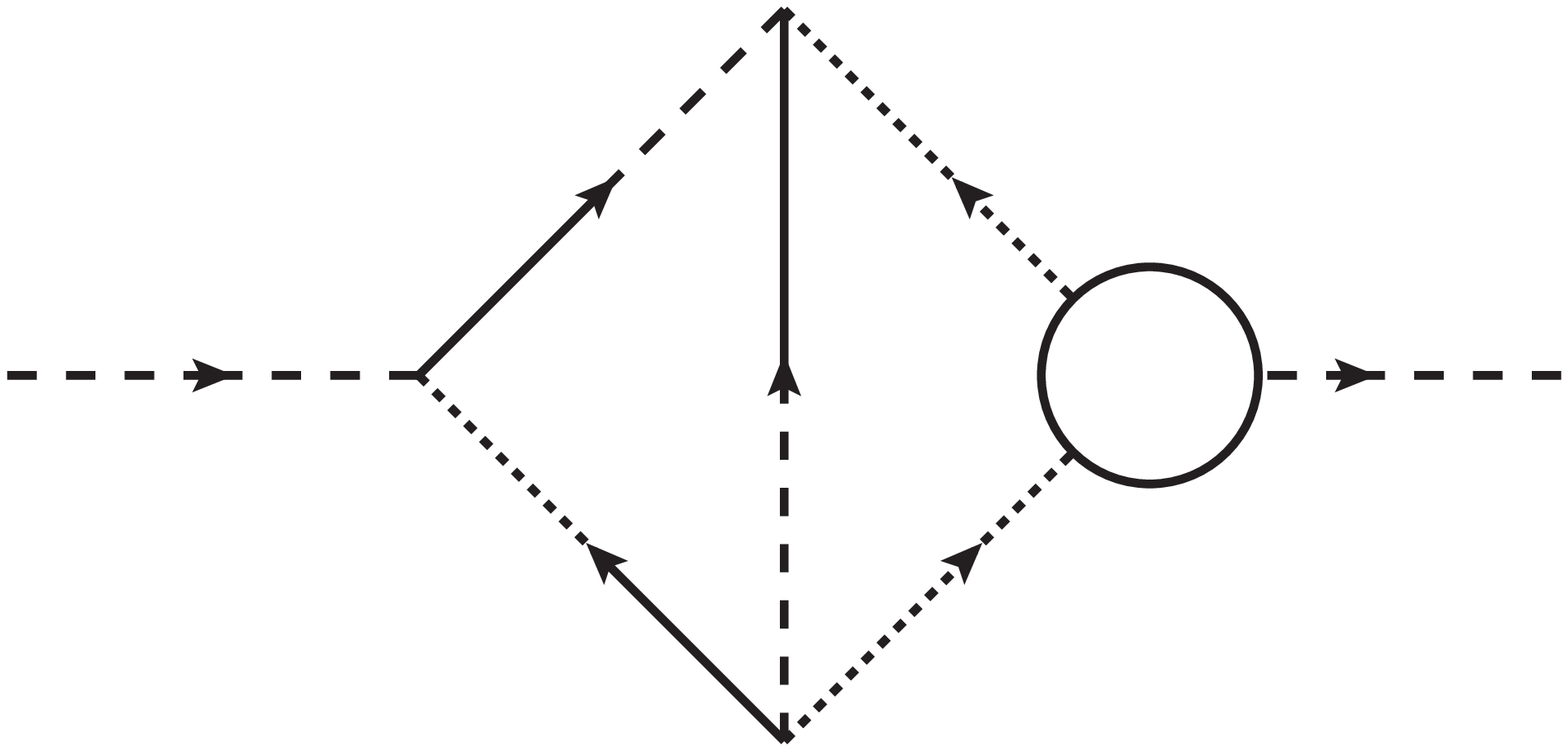}%
 \caption{Graph $ 1\widetilde {B}(r, 0i0)$ }
  \end{center}
  \end{figure}

 \begin{figure}
 \begin{center}
  \includegraphics[width=7.5cm]{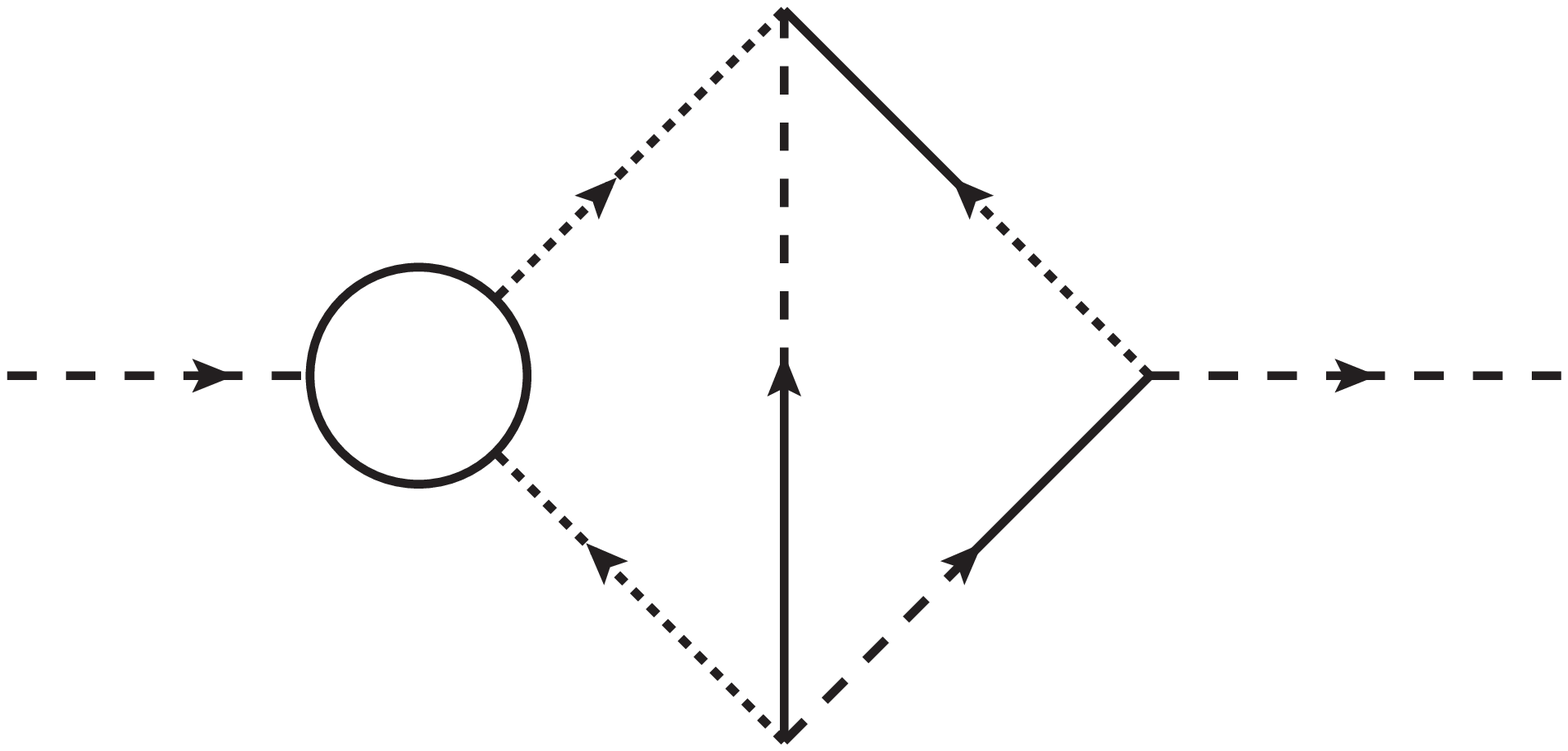}%
 \caption{Graph $ 1\widetilde {B}(l, 0i0)$ }
 \end{center}
 \end{figure}

 The sum of these two graphs is

 \begin{equation}
 \label{57}
 Set 9=\alpha_{ab}\{ {1\over{p_0r'_0}}P_i[Q_jS(q)+Q'_jS(q')]+{1\over{q_0r'_0}}Q_j[P_iS(p)+P'_iS(p')]\}.
 \end{equation}

 \subsection{Set 10}

 We have cancelled linear energy divergences in the first nine sets of graphs. Using changes of variables of integration
 $ p\leftrightarrow -p', q \leftrightarrow -q', r \leftrightarrow -r, k \leftrightarrow k, (i,j)\leftrightarrow (i,j) $
 and $ {{p'_0}\over {p_0}}\approx 1,$ $ {{q'_0}\over{q_0}}\approx 1, $ we transform $ S(p) $ and $ S(q) $ terms into
 $ S(p')$ and $ S(q') $ and obtain for the sum of the graphs so far

 \begin{eqnarray}
 \label{58}
  Set 1+ Set 2+...+Set 9 ~~~~~~~~~~~~~~~~~~~~~~~~~~~~~~~~~~~~~~~~~~~~~~~~~~~~~~ \nonumber \\
   =-\alpha_{ab}
   \{ S(r')[{1\over{p_0q'_0}}P_iQ'_j+{1\over{p'_0q_0}}P'_iQ_j+{1\over{p_0r'_0}}P_i(Q+2Q')_j+{1\over{q_0r'_0}}(P+2P')_iQ_j] && \nonumber\\
    +S(p')[{1\over{q_0r'_0}}(P+P')_iQ_j+{1\over{q'_0r'_0}}P'_iQ'_j] +S(q')[{1\over{p_0r'_0}}P_i(Q+Q')_j+{1\over{p'_0r'_0}}P'_iQ'_j]\}.
  \end{eqnarray}

 But we have two more $ V $-graphs (graphs with fermion loop three-point function) which contain linear energy divergences.
 They are

 \begin{equation}
 \label{59}
 2B(b, 0i0)=-\alpha_{ab}P_iQ'_j\{{1\over{p_0p'_0}}[S(r')-S(q)]+{1\over{p_0q_0}}[S(r')-S(p')]-{1\over{p_0r'_0}}[S(q)+S(p')]\}
 \end{equation}
 and

 \begin{figure}
 \begin{center}
  \includegraphics[width=7.5cm]{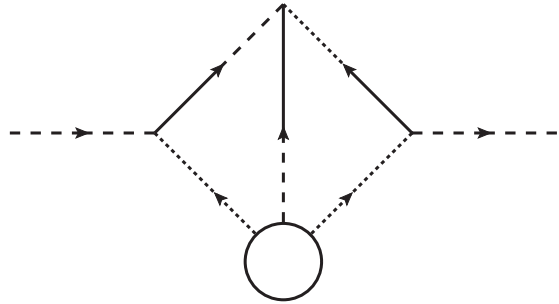}%
 \caption{Graph $ 2B(b, 0i0) $ }
 \end{center}
 \end{figure}

\begin{figure}
 \begin{center}
  \includegraphics[width=7.5cm]{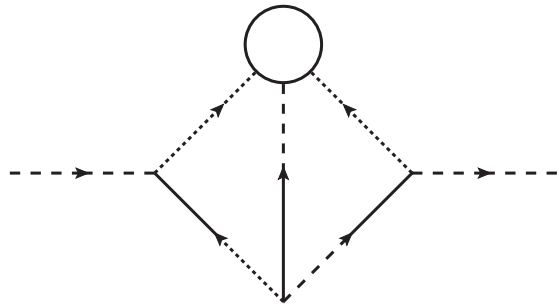}%
 \caption{Graph $2\widetilde {B}(t,0i0) $}
 \end{center}
 \end{figure}
 
 \begin{equation}
 \label{60}
 2\widetilde {B}(t, 0i0)=-\alpha_{ab}P'_iQ_j
   \{ {1\over{q_0q'_0}}[S(r')-S(p)]+{1\over{q_0p_0}}[S(r')-S(q')]-{1\over{q_0r'_0}}[S(p)+S(q')]\}.
 \end{equation}

 Thus the total contribution from vertex part and self-energy insertions is given by the sum of
 (58), (59) and (60).

 The graph where the fermion loop is connected to the rest of the graph with three Coulomb lines (function $ V_{000}$)
 has no $ K^2 $ contribution.

 \subsection { W GRAPHS}

$ W $-graphs contain 4-point fermion loop insertion.
 We start with the sum of two graphs with the $ W_{000l}$ function.

 \begin{eqnarray}
 \label{61}
 W1+W2=-\alpha_{ab}(P+P')_iQ_j\{{1\over{p_0p'_0}}[S(q')-S(r')]+{1\over{q_0q'_0}}[S(p)-S(r')]+{1\over{p'_0q'_0}}[S(q)-S(r')] &&  \nonumber\\
  -{1\over{q_0p'_0}}S(r')+{1\over{q_0(q'+p)_0}}[S(p')-S(q)]+{1\over{q_0p'_0k_0}}[q_0S(q)-q'_0S(q')]\}. 
 \end{eqnarray}

Next two graphs are obtained from $W1$ and $W2$ by rotating them about the vertical axis and the change of variables of integration
$p\leftrightarrow -p', q\leftrightarrow -q', k\leftrightarrow k, (i,j \leftrightarrow i,j) $ 
(so we do not draw them explicitly).

\begin{eqnarray}
\label{62}
\widetilde {W}1+\widetilde {W}2=-\alpha_{ab}P'_i(Q+Q')_j\{{1\over{q_0q'_0}}[S(p)-S(r')]+{1\over{p_0p'_0}}[S(q')-S(r')]
 +{1\over{q_0p_0}}[S(p')-S(r')] && \nonumber\\
 -{1\over{p'_0q_0}}S(r')+{1\over{p'_0(p+q')_0}}[S(q)-S(p')]+{1\over{p'_0q_0k_0}}[p_0S(p)-p'_0S(p')]\}
\end{eqnarray}

Two following graphs contain fermion loop insertion connected to the rest of the graph with two Coulomb and two transverse lines.

\begin{equation}
\label{63}
W3=\alpha_{ab}P_iQ_j\{ {1\over{p_0p'_0}}[S(q)-S(r')]+{1\over{q_0q'_0}}[S(p)-S(r')]-{2\over{p_0q_0}}S(r')\}
\end{equation}

\begin{figure}
\begin{center}
 \includegraphics[width=7.5cm]{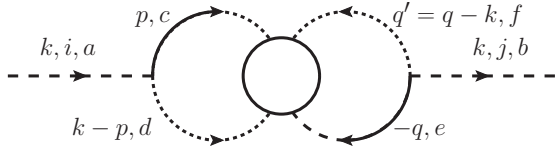}%
 \caption{Graph $W1$. Fermion loop insertion is connected with three Coulomb and one transverse line to the rest of the graph.}
\end{center}
\end{figure}

\begin{figure}
\begin{center}
 \includegraphics[width=7.5cm]{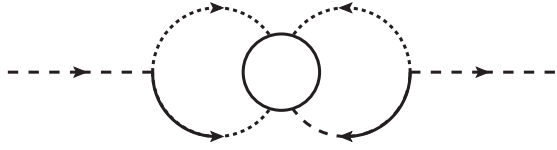}%
 \caption{Graph$W2$}
 \end{center}
 \end{figure}

\begin{figure}
\begin{center}
 \includegraphics[width=7.5cm]{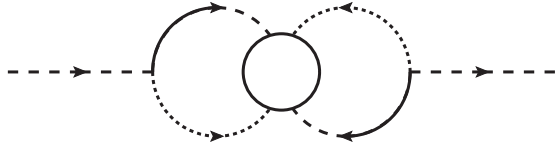}%
 \caption{Graph $W3$}
\end{center}
\end{figure}

\begin{equation}
\label{64}
 W4(K^2)=0
\end{equation}

\begin{figure}
\begin{center}
 \includegraphics[width=7.5cm]{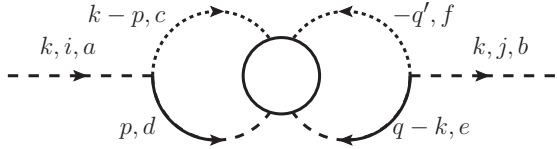}%
\caption{ Graph $W4$. This graph has no $ K^2 $ contribution. }
\end{center}
\end{figure}

 Equations (61) and (62) show new type of quadratic divergence
  in the form ${1\over{k_0}}. $
 These divergences will be cancelled by graphs which contain $ W_{0000}$ insertion.
 There are two such distinct graphs. We list the sum of their contribution.
  
\begin{eqnarray}
\label{65}
W^1_{0000}+W^2_{0000}=-{1\over 2}\alpha_{ab}(P+P')_i(Q+Q')_j\{{1\over{p_0q_0}}[S(k)+S(r')-S(q')-S(p')] && \nonumber\\
                      -{1\over{p'_0q'_0}}[S(k)+S(r')-S(p)-S(q)] 
                      +[{1\over{q_0q'_0}}S(r')+{1\over{q_0k_0}}S(p')-{1\over{q'_0k_0}}S(p)] && \nonumber\\
                      +[{1\over{p_0p'_0}}S(r')+{1\over{p_0k_0}}S(q')-{1\over{p'_0k_0}}S(q)]\}
\end{eqnarray}

\begin{figure}
\begin{center}
 \includegraphics[width=7.5cm]{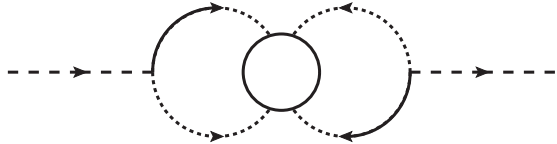}%
\caption{Graph $W^1_{0000}.$ Four Culomb lines attach to the fermion loop.
         The graph contains divergences of the form ${1\over{k_0}}.$ }
\end{center}
\end{figure}

\begin{figure}
\begin{center}
 \includegraphics[width=7.5cm]{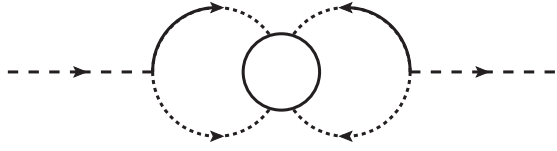}%
\caption{Graph $ W^2_{0000}.$ The graph shows divergences of the form ${1\over{k_0}}.$ }
\end{center}
\end{figure}

The tricky terms with ${1\over{k_0}}$ in (65) we denote as

\begin{eqnarray}
\label{66}
 X_p={1\over{q_0k_0}}S(p')-{1\over{q'_0k_0}}S(p), && \nonumber\\
 X_q={1\over{p_0k_0}}S(q')-{1\over{p'_0k_0}}S(q).
\end{eqnarray}

Then each of $X_p$ and $X_q$ is invariant under $ p,q \rightarrow -p', -q'.$
Therefore, by making these changes of variables, we have

\begin{equation}
\label{67}
-{1\over 2}\alpha_{ab} (P+P')_i(Q+Q')_j[X_p+X_q]=-\alpha_{ab}P'_i(Q+Q')_jX_p-\alpha_{ab}(P+P')_iQ_jX_q.
\end{equation}

First we prove the cancellation of the tricky ${1\over{k_0}}$ divergences. Such terms exist in (61), (62) and (67).
The sum of the last terms in (61) and (62) with (67) is

\begin{eqnarray}
\label{68}
-\alpha_{ab}(P+P')_iQ_j\{{1\over{p'_0k_0}}[S(q)-S(q')]+{1\over{q_0p'_0}}S(q')+{1\over{p_0k_0}}S(q')-{1\over{p'_0k_0}}S(q)\} && \nonumber\\
-\alpha_{ab}P'_i(Q+Q')_j\{{1\over{q_0k_0}}[S(p)-S(p')]+{1\over{p'_0q_0}}S(p)+{1\over{q_0k_0}}S(p')-{1\over{q'_0k_0}}S(p)\} && \nonumber\\
=\alpha_{ab}(P+P')_iQ_j\{ {1\over{p_0p'_0}}S(q')-{1\over{q_0p'_0}}S(q')\}+\alpha_{ab}P'_i(Q+Q')_j\{{1\over{q_0q'_0}}S(p)
 -{1\over{p'_0q_0}}S(p)\}
\end{eqnarray}

The important point to notice is that after cancellation of ${1\over{k_0}} $ divergences, we are left with linear energy divergences
in (68). Now we are ready to prove the cancellation of linear divergences. Linear divergences (i.e. terms with ${1\over{p_0p'_0}}$
and ${1\over{q_0q'_0}}$) are parts of (65), (61), (62), (63), (59) and (60).

 From (65) taking into account (68) we have
\begin{eqnarray}
\label{69}
 A^{ab}_{ij}=-{1\over 2}\alpha_{ab}(P+P')_i(Q+Q')_j[{1\over{q_0q'_0}}S(r')+{1\over{p_0p'_0}}S(r')] && \nonumber\\
+\alpha_{ab}[(P+P')_iQ_j{1\over{p_0p'_0}}S(q')+P'_i(Q+Q')_j{1\over{q_0q'_0}}S(p)] && \nonumber\\
=\alpha_{ab}\{(P+P')_iQ_j{1\over{p_0p'_0}}[S(q')-S(r')]+P'_i(Q+Q')_j{1\over{q_0q'_0}}[S(p)-S(r')]\}
\end{eqnarray}

The linearly divergent part of (61) is
\begin{equation}
\label{70}
 B^{ab}_{ij}=-\alpha_{ab}(P+P')_iQ_j\{{1\over{p_0p'_0}}[S(q')-S(r')]+{1\over{q_0q'_0}}[S(p)-S(r')]\}
\end{equation}

In (62) the linearly divergent part is
\begin{equation}
\label{71}
 C^{ab}_{ij}=-\alpha_{ab}P'_i(Q+Q')_j\{{1\over{p_0p'_0}}[S(q')-S(r')]+{1\over{q_0q'_0}}[S(p)-S(r')]\}
\end{equation}

From $W3$ in (63),

\begin{equation}
\label{72}
D^{ab}_{ij}=\alpha_{ab}P_iQ_j\{ {1\over{p_0p'_0}}[S(q)-S(r')]+{1\over{q_0q'_0}}[S(p)-S(r')]\}.
\end{equation}

Graph $2B(b, 0i0)$ in (59) contributes

\begin{equation}
\label{73}
E^{ab}_{ij}=-\alpha_{ab}P_iQ'_j{1\over{p_0p'_0}}[S(r')-S(q)]
\end{equation}

and graph $2\widetilde {B}(t, 0i0)$ in (60) gives

\begin{equation}
\label{74}
F^{ab}_{ij}=-\alpha_{ab}P'_iQ_j{1\over{q_0q'_0}}[S(r')-S(p)].
\end{equation}

It is easy to check that the sum of equations (69) to (74) gives zero (using the symmetry $p\rightarrow -p', q\rightarrow -q',
(i,j) \rightarrow (i,j) $).

\subsection {\it {A-divergences}}
Having verified that all the linear energy divergences cancel, we are ready to come to the
main point of the paper, the cancellation or otherwise of the {\it A- ambiguous} integrals like (3).
We collect the remaining {\it A- divergences}.
From (65) we have,

\begin{eqnarray}
\label{75}
\widetilde{A}^{ab}_{ij}=-{1\over 2}\alpha_{ab}(P+P')_i(Q+Q')_j\{ {1\over{p_0q_0}}[S(r')-S(q')-S(p')]-{1\over{p'_0q'_0}}[S(r')-S(p)-S(q)]\}
 && \nonumber\\ -\alpha_{ab}\{(P+P')_iQ_j{1\over{q_0p'_0}}S(q')+P'_i(Q+Q')_j{1\over{q_0p'_0}}S(p)\} && \nonumber\\
=-\alpha_{ab}{1\over{q_0p'_0}}[(P+P')_iQ_jS(q')+P'_i(Q+Q')_jS(p)].
\end{eqnarray}

The remaining {\it A-divergences} in (61) are

\begin{equation}
\label{76}
\widetilde {B}^{ab}_{ij}=-\alpha_{ab}(P+P')_iQ_j\{{1\over{q'_0p'_0}}[S(q)-S(r')]-{1\over{p'_0q_0}}S(r')-{1\over{q_0r'_0}}[S(p')-S(q)]\}.
\end{equation}

From (62) we have the {\it A-divergences}

\begin{equation}
\label{77}
\widetilde {C}^{ab}_{ij}=-\alpha_{ab}P'_i(Q+Q')_j\{{1\over{q_0p_0}}[S(p')-S(r')]-{1\over{p'_0q_0}}S(r')-{1\over{p'_0r'_0}}[S(q)-S(p')]\}.
\end{equation}
The {\it A- divergence} in (63) is

\begin{equation}
\label{78}
\widetilde {D}^{ab}_{ij}=-\alpha_{ab}{2\over{p_0q_0}}P_iQ_jS(r').
\end{equation}

In (59) we have

\begin{equation}
\label{79}
\widetilde {E}^{ab}_{ij}=-\alpha_{ab}P_iQ'_j\{{1\over{p_0q_0}}[S(r')-S(p')]-{1\over{p_0r'_0}}[S(q)+S(p')]\}.
\end{equation}

In (60) the remaining divergence is

\begin{equation}
\label{80}
\widetilde {F}^{ab}_{ij}=-\alpha_{ab}P'_iQ_j\{{1\over{q_0p_0}}[S(r')-S(q')]-{1\over{q_0r'_0}}[S(p)+S(q')]\}
\end{equation}

Summing up (75) to (80) with (58)(using the same changes of variables to transform $S(p)$ and $S(q)$ terms into
$S(p')$ and $S(q')$) and the rule of factorization (5), the final result for the sum of all the {\it A- divergences} is
\begin{eqnarray}
\label{81}
X^{ab}_{ij}=\alpha_{ab}\{S(r')[P_iQ_j({1\over{p'_0q_0}}+{1\over{p_0q'_0}}+{1\over{p_0q_0}})+{2\over{p_0q'_0}}P_iQ'_j] && \nonumber\\
  -{1\over{p'_0r'_0}}P'_iQ_jS(p')-{1\over{q'_0r'_0}}P_iQ'_jS(q')\}.
\end{eqnarray}
The {\it A- divergences} do not cancel out. \\

The first line of (81) contains terms like
\begin{equation}
\label{82}
\alpha_{ab}S(r')P_iQ_j{1\over p_0q'_0}
\end{equation}
in which the ${\mu}^2$ subtraction term in (10) gives zero contribution by the factorization rule (5).
But the $(-r'^2)^{-\epsilon}$ term in (10) gives a non-zero contribution to (82) which is
in fact proportional to $\Gamma({{\epsilon}\over 2})$. The second line of (81), using a change of variables,
and the identity
$${1\over{p'_0r'_0}}+{1\over{q_0r'_0}}+{1\over {p'_0q_0}}=0$$
(this identity is to be used only for large energies, it is consistent with (4) being convergent)
and using the factorization rule, may be written
\begin{equation}
\label{83}
\alpha_{ab}P'_iQ_j{1\over q_0r'_0}[S(p')-S(q)].
\end{equation}
Again the ${\mu}^2$ subtraction term in (10) cancels out, but the remaining two parts of (83)
give different non-zero contributions, each proportional to $\Gamma({{\epsilon}\over 2})$.
Thus, in (81),  the UV divergences and the {\it A-integrals} conspire to give a divergent result.\\

\section {Conclusion}

To two loop order, the non-convergent {\it A-integrals} of the form (3) are rendered harmless
by the use of the identity (4). But to three loops, when there are UV divergent sub-graphs,
we find that a similar process fails to work. So we conclude that it is not possible
to make sense of Coulomb gauge perturbation theory to three loop order.\\

Acknowledgements \\
This work was supported by the Ministry of Science and Technology of the Republic of Croatia under contract No. 098-0000000-2865(A.A.).
We are grateful to Dr. G. Duplan\v ci\'c for drawing the figures.


\begin{thebibliography}{10}
  
  \bibitem{Cucchieri} A. Cucchieri, hep-lat/0612004, AIP Conf. Proc. {\bf 892},22-28 (2007)
  
  \bibitem{AT} A. Andra\v si, J.C. Taylor, Eur. Phys. J. C {\bf 41},377 (2005)

  \bibitem{AT2} A. Andra\v si, J.C. Taylor, Ann. of Phys. {\bf 324},No.10, 2179-2195 (2009)

  
  \bibitem{mohapatra} R.N. Mohapatra, Phys. Rev. D4 {\bf 22},378, 1007 (1971)
  
  \bibitem{cheng} H. Cheng, E. C. Tsai, Phys. Lett. B {\bf 176} 130 (1986);
                  Phys. Rev. Lett. {\bf 57} 511 (1986)
  
  \bibitem{doust} P. Doust, Ann. of Phys. {\bf 177}, 169 (1987)
  
  \bibitem{christlee} N. Christ, T. D. Lee, Phys. Rev. D{\bf 22}, 939 (1980)
  
  \bibitem{taylor} J. C. Taylor, in Physical and Nonstandard Gauges, Proceedings,
                   Vienna, Austria 1989, edited by P. Gaigg, W. Kummer, M. Schweda
  \end{thebibliography}
  \end{document}